\documentclass[twocolumn,showpacs,preprintnumbers,amsmath,amssymb]{revtex4}
\usepackage{epsfig}
\newcommand{\bm}[1]{ \mbox{\boldmath $#1$}  }

\begin{document}

\title{Distinction between sequential and direct three-body decays }

\author{R. \'Alvarez-Rodr\'{\i}guez, H.O.U. Fynbo, A.S. Jensen, }
\affiliation{ Department of Physics and Astronomy, University of Aarhus, 
DK-8000 Aarhus C, Denmark}
\author{E. Garrido }
\affiliation{Instituto de Estructura de la Materia, CSIC, Serrano 123, 
E-28006 Madrid, Spain}

\date{\today}

\begin{abstract}
We discuss the three-body decay mechanisms of many-body resonances.
Sequential decays proceed via two-body configurations after emission
of the third particle.  In direct decay all three particles leave
simultaneously their interaction regions.  The intermediate paths
within the interaction regions are not observables and only accessible
through models.  The momentum distributions carry, apart from
polarization, all possible information about decay modes and resonance
structure. In this context we discuss detailed results for the decay of
the $^{9}$Be($5/2^-$) resonance. 
\end{abstract}


\maketitle

\paragraph*{Introduction.}
 
The distinction between different decay mechanisms of a given
resonance is a general problem in quantum mechanics
\cite{dal53,ams98,gri05}.  The properties of the fragments emerging after
the decay carry all possible information. A number of accurate
and kinematically complete measurements appear these years in the
literature for three-body decays \cite{gal05,bla03,fyn03}.  The
results are often used to interpret the decay as either sequential
via an intermediate configuration or direct to the continuum
\cite{pap07}.  Also theoretical models are used to classify the decay
mechanisms \cite{gar07,alv07b} but the results of interpretations from
different groups of comparable data are not always in agreement with
each other \cite{gri05,pap07,des01}.  This is unfortunate since a
proper classification of the different decay modes is used both in
analysis of data and in subsequent applications.

An important question is then to which extent are these disagreements
real or only related to different models and choices of basis.  A
correct understanding is of crucial importance for several reasons,
e.g.

(i) The analyses of data are based on an $R$-matrix formalism which
intrinsically assumes sequential decay via different intermediate
configurations. No other procedure to analyze the data exists.
Improvements are in addition at best included to account for
symmetries and the interaction between the first emitted particle and
the remaining subsystem.  

(ii) The intermediate ``paths'' leading from initial to final states
are in quantum mechanics not observables.  The measured final state
information only relates to the initial state through the filter of
intermediate configurations.  Initial state information and
classifications into decay modes are then necessarily the results of
interpretations or model computations.  Understanding the underlying
physics picture is then necessary for predictions.

(iii) The interpretations are used e.g. to derive the reaction rates
for the inverse process of absorption in astrophysical environments.
Reliable applications must use the correct rates which may depend on
the adopted interpretation of the decay data.

The purpose of the present letter is to investigate the distinct
features of different decay modes.  We illustrate by comparing
computed and experimental results and conflicting interpretations for
decay of $^{9}$Be($5/2^-$).

\paragraph*{Procedure.}

To describe three-body decay of a many-body resonance we
must use a three-body model.  Effects of many-body character at
short-distances are mocked up by use of a structureless short-range
phenomenological three-body potential which also is fine-tuning 
the resonance energy.  The three particles have masses $m_i$ and coordinates
$\bm{r_{i}}$, $i=1,2,3$.  To compute the resonance wave functions we
use the Faddeev equations and the hyperspherical expansion method
\cite{nie01} combined with complex scaling \cite{fed02}.  
The crucial coordinate is the hyperradius $\rho$ defined as a mass
averaged distance, i.e.  $m \rho^2 = \sum_{i<j} m_i m_j (\bm{r_{i}} -
\bm{r_{j}})^2/M $, where $M=m_i+m_k+m_j$ and $m$ is an arbitrary
normalization mass chosen to be the nucleon mass.  The remaining five
intrinsic coordinates are dimensionless angles.  We apply the
transformation $\rho \rightarrow \rho \exp(i\theta)$ for a given
scaling angle $\theta$ and compute the adiabatic potentials.  When
$\theta > \theta_{2} = 0.5 \arctan(\Gamma_2/2E_2)$ where $E_2$ is the
energy and $\Gamma_2$ is the width, one of the potentials describes
asymptotically this two-body resonance with the third particle far
away.

To illustrate we compare measured distributions \cite{pap07} and
computations for the $5/2^-$-resonance of $^{9}$Be ($\alpha + \alpha +
$ neutron).  The two-body interactions reproduce the low-energy
$\alpha-\alpha$ and $n-\alpha$ scattering data \cite{alv07,gar02}.
The three-body interaction is used to reproduce the measured resonance
energy above the three-particle breakup threshold of $0.856$~MeV$-i
0.4$~keV (excitation energy 2.429~MeV).

\paragraph*{Adiabatic potentials.}

\begin{figure}[h]
\begin{center}
\vspace*{-1.2cm}
\epsfig{file=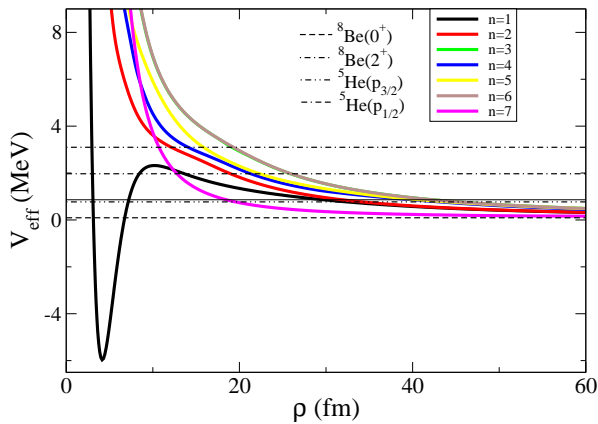,scale=0.32,angle=270}
\end{center}
\vspace*{-0.9cm}
\caption{The real parts for $\theta = 0.1$ of the lowest potentials, 
including the three-body potential, for the $0.856$~MeV
$^{9}$Be($5/2^{-}$)-resonance (horizontal full line) as function of $\rho$. }
\label{fig1}
\end{figure}

The real parts of the coupled set of adiabatic potentials are shown in
fig.~\ref{fig1}.  The potential $n=1$ has a pocket at small
distance holding the resonance at the measured energy.  All the other
potentials are repulsive for all distances. The potential $n=7$
becomes the lowest at around $\rho \approx 12$~fm and approaches the
$^{8}$Be($0^+$) resonance energy (dashed line) with a corresponding
structure combined with the neutron far away.  The two-body resonance
energies of $^{8}$Be($2^+$) and $^{5}$He($1/2^-$) are above the first
potential for almost all distances while the energy of
$^{5}$He($3/2^-$) only is above the first potential when $\rho$ is
larger than $30$~fm.  The angles $\theta_2$ for these three resonances
are all larger than $\theta = 0.1$.

The population of the adiabatic components varies rapidly at small
distances where the coupled adiabatic potentials move relative to each
other.  The couplings, when levels get close, are decisive for the
choice between maintaining the structure or following the
energetically most favorable path.  In the present case 
about 96\% of the wave function at large
distance maintains the structure related to the pocket at small
distance.  About 3\% is located in the component related to the ground
state structure of $^{8}$Be.  The relative populations become
essentially constant at large distance although the $^{8}$Be($0^+$)
description is less accurate.

\begin{figure}[h]
\begin{center}
\vspace*{-1.6cm}
\epsfig{file=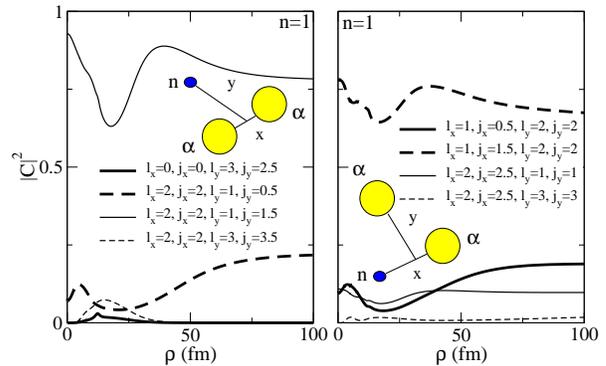,scale=0.32,angle=270}
\end{center}
\vspace*{-1.1cm}
\caption{The probabilities $|C|^2$ for different angular momentum
combinations ($l_x,j_x,l_y,j_y$) in the two Jacobi coordinates for
the dominating (96\%) adiabatic component for the
$^{9}$Be($5/2^{-}$)-resonance. }
\label{fig2}
\end{figure}

{\em The partial wave decomposition} of the dominating component
is shown in fig.~\ref{fig2}. The angular momentum combinations at
large distance are almost entirely the $^{8}$Be($2^+$) structure of
$2\hbar$ between the two
$\alpha$-particles coupled to $1\hbar$ between their center-of-mass
and the neutron (fig.~\ref{fig2}, left). The same wave function also 
has the $^{5}$He-structure of $1\hbar$ between the neutron and one 
$\alpha$-particle coupled to $2\hbar$ between their center-of-mass 
and the other $\alpha$-particle (fig.~\ref{fig2}, right).

\paragraph*{Energy distributions.}

\begin{figure}[h]
\begin{center}
\vspace*{-1.4cm}
\epsfig{file=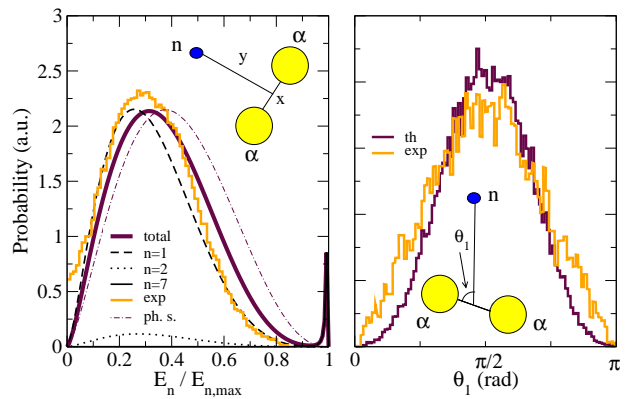,scale=0.34,angle=270}
\end{center}
\vspace*{-1.2cm}
\caption{The measured \cite{pap07} and computed normalized distributions of 
the neutron energy (left) in units of the maximum, and the angle
$\theta_1$ (right) between the relative momenta of the
$\alpha$-particles and the neutron momentum.  Phase
space is abbreviated ph.s.  }
\label{fig3}
\end{figure}

The computed neutron energy distribution (left in fig.~\ref{fig3})
almost reproduces the experimental curve, although 
lies below at small energy and
above at high energy. However the data is obviously an overestimate 
at small energy, and due to the normalization this
appears as an underestimate at high energy.
The computed small peak at almost
maximum neutron energy arises from the potential, 
corresponding to decay via the ground state of $^{8}$Be.  Essentially
all other parts of the broad and smooth distribution arise from the
$n=1$ potential which in no way corresponds to any two-body
structures.  In fact the distribution solely determined from phase
space for the largest angular momentum component in fig.~\ref{fig2} only
should be shifted a little to match the computed results. The angular
distribution in fig.~\ref{fig3} between the direction of the neutron
and the relative momentum of the two $\alpha$-particles also compares
favorably with the measurement.

The traditional analysis in terms of tails of broad higher-lying
resonances was applied in the recent experimental investigation
\cite{pap07} where the $^{8}$Be($2^+$) path was claimed to account for
about 86\% of the decay.  This is in agreement with the theoretical
computation in \cite{des01} but in contrast to \cite{gri05} where
$^{8}$Be($2^+$) was ignored and the $^{5}$He-channel was found to
contribute a little on top of a ``democratic'' ($\sim 94\%$)
(phase-space determined) decay.  In \cite{pap07} it is also claimed that
the decay via $^{5}$He($p_{3/2}$) cannot explain the data as well as the
$^{8}$Be($2^+$) channel.  This conclusion is based on analysis of the
angular distribution while the energy distribution is equally well
explained via both channels.  This difference can be attributed to the
neglect of $^{5}$He($p_{1/2}$) which with $^{5}$He($p_{3/2}$) is
needed to be equivalent to $^{8}$Be($2^+$), see fig.~\ref{fig2}.

\paragraph*{Spatial structure}

\begin{figure}[h]
\vspace*{-0.9cm}
\hspace*{-0.7cm}
\epsfig{file=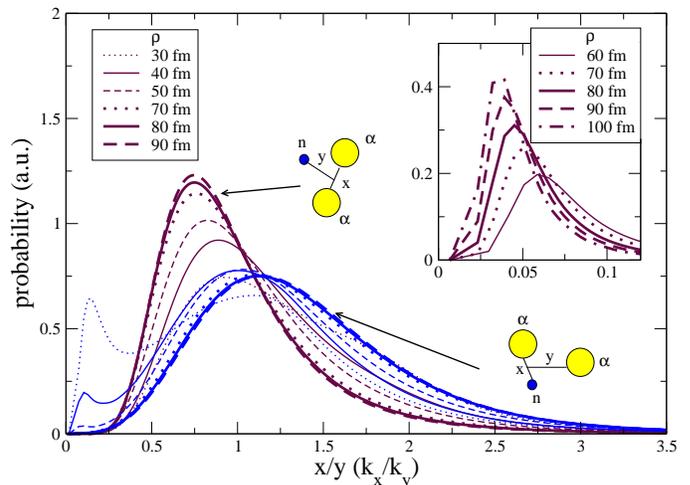,scale=0.37,angle=270}
\vspace*{-0.98cm}
\caption{The evolution with hyperradius for two sets of Jacobi coordinates
of the probability distributions for the ratios between distances or
momenta of two particles and their center-of-mass and the third
particle.  The inset shows the contribution from $n=7$ for $^{8}$Be. }
\label{fig4}
\end{figure}

The decaying resonance wave function depends on the relative
coordinates of the three particles.  The variation describes the
dynamical evolution of the substructures from before the barrier, via
intermediate configurations to large distances. The structure of the
resonance wave function can be seen in fig.~\ref{fig4} where we
compare the probability for finding given values of the ratio of the
distance between two particles and the distance from their
center-of-mass to the third particle.

The peak in the first coordinate system ($^{8}$Be) stabilizes at
around 0.75 for $\rho$ larger than about $60$~fm.  This means that when
the system emerges from underneath the barrier at $\rho = 30$~fm, see
fig.~\ref{fig1}, the most probable distance between
the two $\alpha$-particles is about 18~fm 
which directly shows that the spatial $^{8}$Be-structure is not
present.  In contrast the peak in the inset moves towards smaller
values as $\rho$ increases.  This peak arises entirely from the lowest
potential at large $\rho$, see fig.~\ref{fig1}, with the two
$\alpha$-particles in a relative $0^+$-structure and a distance of
about 3.3~fm independent of $\rho$.  This reflects that
$^{8}$Be($0^+$) is populated and the $\alpha$-particles moves apart as
for a decaying two-body resonance.

For the small $\rho$-values, the other coordinate system ($^{5}$He)
reveals a small peak at small values connected to a rather broad
distribution for higher ratios. The distance between the neutron and
one $\alpha$-particle is about 2.8~fm for this small peak while the
distance to the other $\alpha$-particle is about 27~fm.  Combined with
the partial wave decomposition in fig.~\ref{fig2} this means a small
probability for a structure resembling $^{5}$He.  However, as $\rho$
increases this peak disappears before the couplings between different
potentials have vanished.  The distribution stabilizes as the broad
peak around 1.2 with two $\alpha$-particles each with comparable
relative distances to the neutron.

\paragraph*{Sequential and direct decays.}

The observables are related to the asymptotic large-distance results.
We classify the possible decay modes as direct to the three-body
continuum or sequential via two-body configurations.  The different
decay modes can be found from the rotated wave function if the scaling
angle is larger than the angles of the relevant two-body resonances.

For direct decays all three particles simultaneously leave the
interaction region.  This corresponds to converged distributions at
the largest accurately computed $\rho$-values.  Further increase of
$\rho$ increases all distances proportional to $\rho$, and the points
in fig.~\ref{fig4} can be translated into relative distances between
the particles.  Spatial and momentum distributions are asymptotically
identical since the particles at large distances move along their
respective coordinates relative to the center-of-mass.

The decay is sequential if the conserved two-body quantum numbers
match an intermediate two-body structure and the corresponding
distance in the complex scaled wave function remains small and
constant until all energy exchanges, or couplings, with the third
particle can be neglected.  Each sequential decay corresponds to
asymptotic population of the adiabatic component describing that
structure.  These decays result in distributions which continue to
change with increasing $\rho$ like in the inset of fig.~\ref{fig4}.
In at least one of the three possible Jacobi coordinates the
distribution is moving towards small values of the corresponding
distance ratio.  Then two particles remain close in the intermediate
structure while the increase of the average size is achieved entirely
by the third particle moving away. These components must be removed
from the rotated wave function and treated separately as two
consecutive two-body decays with precisely known kinematics.  They are
decoupled from each other and from the direct decay components.

One two-body configuration could have attraction enough to favor an
intermediate resonance structure to (exceedingly) large average
distances when one particle is far outside normal interaction ranges.
This could happen when the total energy is smaller than the energy of
the two-body resonance and the effective barrier in fig.~\ref{fig1}
becomes very wide. The system remains in the classically forbidden
region to large distances. The signature is stable population at
intermediate $\rho$ of one adiabatic component.  At larger $\rho$ the
population necessarily decreases due to energy conservation
corresponding to decay directly into the three-body continuum.  The
three-body decay width can be extremely small for such peculiar
structures which have been baptized as virtual sequential decay
\cite{gar04}. This is decay through the tail of an energy forbidden 
low-lying two-body resonance.  The treatment can be either to continue
the computation to much larger $\rho$ or handle separately as for
sequential decay.

These decay modes are in fact different intermediate paths and as such
not observables.  In pronounced cases they leave distinctly different
signatures in computed distributions as shown in fig.~\ref{fig4}, but
otherwise a distinction is not always meaningful and mode separation
becomes artificial or model dependent.

\paragraph*{Scaling angle dependence.}

The presented results are all for $\theta= 0.1$. Increasing $\theta$
to 0.14 and 0.24 beyond the $^{8}$Be($2^+$) and $^{5}$He($p_{3/2}$)
resonances, respectively, do not change the momentum distributions or
the fractions of sequential decay.  However, the accuracy decreases
with increasing $\theta$ which therefore is kept as small as possible.

\paragraph*{Geometric visualization.}

One pertinent question is whether it is possible to distinguish the
different decay modes by use of the experimental information without
theoretical models.  Let us plot the measured 
energy distributions $P(\epsilon_i)$
of all three particles as functions of $\epsilon_i$ defined as the particle
energies divided by the respective maximum values. 
We change $P(\epsilon_i)\:d\epsilon_i$ into the distribution
$G(x/y)\:d(x/y)$ given as function of the variable $x/y$ in fig.~\ref{fig4}.
Since $x/y=m_{ijk}\sqrt{(1-\epsilon_i)/\epsilon_i}$, this defines
$G=2\epsilon_i \sqrt{(1-\epsilon_i)\epsilon_i}\:P(\epsilon_i)/m_{ijk}$, where
$m_{ijk}=(m_j+m_k)\sqrt{m_i/(m_j m_k M)}$.  

The resulting plot of $G(x/y)$ is designed to resemble the direct decays or the
converged part of fig.~\ref{fig4}. However, it contains perhaps also
sequential decay contributions which may exhibit 
characteristic features with a peak at the corresponding energy and a
width equal to the sum of the widths of the intermediate two-body
state and the decaying resonance.  For virtual sequential decay the
peak in the distribution arises from phase space combined with the
tail of the intermediate resonance. 

To visualize the measured decays geometrically, plots of $G$ in
fig.~\ref{fig4} could be constructed.  One could look for
features of two-body kinematics corresponding to sequential
decays. Then it is crucial to correlate consistently the results in
different Jacobi coordinates.  If these features cannot be identified
it is probably because an extraction is ambiguous and the structures
can be described in several basis sets or by different models. Then
the positions of the remaining peaks can be interpreted as length
ratios in triangular configurations.

When two particles are identical only two plots are possible as for
$^{9}$Be.  The two identical particles go
together in an intermediate configuration through $^{9}$Be($0^+$)
and their energy
distributions follow from two-body kinematics.  Everything else is
direct decay, and the $G-$distributions in fig.~\ref{fig4}
correspond to a triangular decay
geometry with side ratios $d_{n\alpha}:d_{n\alpha}:d_{\alpha\alpha} =
1.5:1.3:1$.

For three identical particles, like $^{12}$C-decay into three
$\alpha$-particles, only one such plot exists \cite{alv07b}. Decays of
the two lowest $0^+$-resonances are preferentially sequential through
the ground state of $^{8}$Be resulting in a peak corresponding to
emission of the first particle and the kinematically related broad
structure at lower energy.  The direct decay of the $1^+$-state gives
three peaks and corresponding triangular side ratios of $2.1:1.6:1$.
A directly decaying linear chain would produce a distribution with
peaks at $2/3$ and at a very large value.

\paragraph*{Conclusions.}

The measured energy distributions of three-body decaying resonances
reflect the structure related to the decay mechanism.  Specific plots
are well suited for extraction of the corresponding geometry.  Sharp
peaks most likely correspond to sequential decays via energetically
allowed two-body configurations.  The energy forbidden virtual
sequential decays leave a signature of the corresponding two-body
angular momentum and parity.  Direct decays only carry information of
conserved quantum numbers of the decaying state.  Direct and
sequential decays can be distinguished if energy or angular momentum
signatures are different.  

In general, decays via high-lying or broad resonances are very
unlikely because the corresponding components would be depopulated
already at small hyperradii where the coupling is strong.  Sequential
decays via energy allowed low-lying and narrow resonances are possible
but not unavoidable.  Virtual (energy forbidden) sequential decays are
possible when the two-body resonance energy and width both are small
and the effective barrier therefore very thick.  Several descriptions
in terms of different basis sets may be equally efficient.

For application on the three-body decay of $^{9}$Be($5/2^-$) we find a
small fraction emerging via $^{8}$Be($0^+$) but no decays via the
$^{8}$Be($2^+$) or $^{5}$He($p$) structures.  In this case reaction
rates based on derived branching ratios via sequential channels are
not meaningful or  misleading.

\paragraph*{Acknowledgments.} 

R.A.R. acknowledges support by a post-doctoral fellowship from
Ministerio de Educaci\'on y Ciencia (Spain).  We are grateful for many
discussions with D.V. Fedorov and we thank O. Kirsebom for detecting
several mistakes related to fig.~\ref{fig4}.

\end{document}